# FEATURES OF LOW TEMPERATURE TUNNEL MAGNETORESISTANCE IN PRESSED POWDER OF CHROMIUM DIOXIDE

(DER TIEFTEMPERATUR TUNNEL-MAGNETWIDERSTAND IN GEPRESSTEM CHROMDIOXID)

V.A. Horielyi*, N.V. Dalakova, E.Yu. Beliayev

*B.I. Verkin Institute for Low Temperature Physics and Engineering, NAS of Ukraine*

*47 Lenin's Prospect, 61103, Kharkov, Ukraine.*

E-mail: beliayev@ilt.kharkov.ua

**Abstract**

Resistive and magnetoresistive properties of two samples of compacted powders of ferromagnetic half-metal $CrO_2$ with shape anisotropy of nanoparticles were studied. The powders were prepared by the method of hydrothermal synthesis and consisted of needle-like particles with average diameters ~ 24÷34 nm and mean length of ~ 300 nm. One of the samples has been made of compacted powder of pure $CrO_2$, while another sample has been prepared from a substitution solid solution $Cr_{1-x}Fe_xO_2$. The aim of this work was to study the effect of Fe impurity on the value of tunnel resistance and tunnel magnetoresistance for compacted $CrO_2$ powders. It was found that the addition of Fe impurity leads to an increase in the coercive force of the powder and reduce the tunnel magnetoresistance. We assume resonant tunneling mechanism on the Fe impurities. We found a strong dependence of the magnetoresistance on the spin relaxation rate in the process of magnetization reversal. The possible reasons for such dependence are discussed.

**Introduction**

Chromium dioxide ($CrO_2$) as a ferromagnetic half metal with a Curie temperature $T_C$ ≈ 390 K [1,2] is of great interest for the study of spin-polarized transport and for application in spintronic devices. It is known that the efficiency of spintronic devices depends on the value of maximum achievable magnetoresistance (MR). The intrinsic MR of monocrystalline $CrO_2$ at room temperature is about 1% in a field $H$ = 1 T [3], and varies slightly with decreasing temperature. At the same time, in the composite granular material consisting of $CrO_2$ particles, coated with a thin layer of dielectric, the MR is huge, reaching over 30% at low temperatures and low fields [2,4]. This MR is extrinsic. It is due to the granular structure and depends on the properties of magnetic tunnel junctions between the ferromagnetic grains and the relative orientation of the magnetization vector in the neighboring grains. MR of this kind is called a tunnel magnetoresistance (TMR). The prospects for using of chromium dioxide powder are associated with the development of technology for production of powders with high TMR. In our earlier work [5-7] we have considered the influence of various factors that can be controlled by the technology of synthesis on the magnitude of TMR. These factors include the thickness and the type of dielectric layers between $CrO_2$ particles, and the anisotropy of the particles' shape. The maximum value of negative TMR (about 36% at temperatures $T \leq 5$ K in fields $H$ = 0.3 T) was obtained for the compacted powder sample made of acicular particles with the thickness of dielectric coating ~ 2 nm. This work is aimed at studying the effect of impurities of ferromagnetic metal on the magnitude of TMR in compacted powders of $CrO_2$.

EXPERIMENT

We investigated the resistive and magnetoresistive properties of two compressed powders of chromium dioxide. One of the samples (Sample 1) made of pure $CrO_2$, while the second one (Sample 2) made of $CrO_2$ - Fe solid solution with Fe content 75 mmol / 1 mole of chromium. Both powders were prepared under the same conditions and consisted of particles of prismatic shape with the ratio of particle diameter to its length ~ 1:10. The main parameters for the investigated powders are given in Table. As seen from the Table, introduction of Fe impurities changes all the magnetic characteristics and the lattice parameters of $CrO_2$.

The powders have been prepared by the method of hydrothermal synthesis. Common features of synthesis technology are described in [8]. Particles have been coated by naturally degraded layer consisting of a mixture of amorphous β-CrOOH and some quantity of occluded chromic acid. In Sample 2 iron was presented in two forms: as a solid solution $Cr_{1-x}Fe_xO_2$ and as $Cr_{2-2x}Fe_{2x}O_3$. The first compound provides a high coercive force. The second one is the ballast. It exists in the form of small individual particles. These particles have a magnetization of two orders of magnitude lower than $CrO_2$. $Cr_{2-2x}Fe_{2x}O_3$ does not contribute to the conductivity. Iron concentration varies over the thickness of the particle. On the surface, it is significantly higher and so the magnetic reversal is controlled by the surface composition. By the method of Mössbauer spectroscopy at $^{57}Fe$ atoms in paper [9] was found that $Fe^{3+}$ ions in the chromium dioxide powder are distributed between the three magnetic solid solutions. In addition to the solid solutions $Cr_{1-x}Fe_xO_2$ (this is massive material and iron-enriched surface layer) and particle inclusions $Cr_{2-2x}Fe_{2x}O_3$ iron penetrates in the oxy-hydroxide of chromium β-CrOOH, which is a part of the dielectric membrane.

The temperature dependence of resistivity for two powders was measured at constant current $J$ = 100 μA in the regime of implementation of Ohm's law. The dependence of MR on the magnetic field, defined as $\Delta R(H)/R(0) = [R(H)-R(0)]/R(0)$ was recorded in accordance with the usual experimental procedure for recording the magnetization hysteresis loops.

RESULTS AND DISCUSSION

Fig. 1 shows the non-metallic resistance behavior ($d\rho/dT < 0$) for both samples. At $T$ < 20 K the dependence of $\rho(T)$ is close to an exponential one that witnessed the tunneling nature of conductance. For a solid solution $Cr_{1-x}Fe_xO_2$ (Sample 2) lower resistivity values can be seen. At the same time, negative MR in powder Sample 2 was less (resistance was higher) than in Sample 1. Since the type, quality and the thickness of dielectric coating of particles in two investigated powders are roughly the same, it can be assumed that the observed changes of the resistance value for Sample 2 at zero (Fig. 1) and a finite magnetic field (Fig. 2) are mainly due to the presence of iron impurity. The data presented in Fig. 1 and Fig. 2 show that the effect of impurities on the conductivity of tunnel system is ambiguous. Impurities in the dielectric layer can facilitate ordinary electron tunneling and decrease the resistance of the system (Fig. 1). On the other hand, these impurities hamper magnetic tunneling, which lowers MR (Fig. 2). Reduction of MR in solid solutions $Cr_{1-x}Fe_xO_2$ can be explained as shown in [9] by the formation of resonance states on Fe ions in the surface dielectric layer of the particles. In addition, MR may decrease due to formation of localized states at the impurities $Cr_{2-2x}Fe_{2x}O_3$, some of which are in the intergranular dielectric medium. In the latter case, we have tunneling through a chain of localized electronic states, which reduces the polarization of the tunneling electrons and cause the lowering of negative magnetoresistance. Such mechanism was considered in [10] for tunnel junctions ferromagnet-insulator-ferromagnet including half metals. Random distribution of impurities can also contribute to the decrease in electron polarization.

In Fig. 2 MR curves at low fields reveal a slight positive MR (see Fig. 2b), for which the position of maximum $\pm H_p$ must comply with the coercive force $H_c$. It can be seen that the introduction of Fe impurities increases $H_p$.

In granular half metals tunneling conductance is determined by two factors: the thickness of the dielectric layers and the mutual orientation of magnetic moments of neighboring granules. For this reason, the hysteretic curves of tunnel MR should reflect the well-known hysteretic behavior of magnetization $M(H)$, namely - a sharp increase in $M(H)$ with increasing field in weak fields, followed by a weak growth to saturation. In this case the MR hysteretic curves show no correlation between the behavior of the $M(H)$ and the behavior $\Delta R(H)$. In a small field $H_X > H_p$ we observe an additional intersection of $\Delta R(H)$ curves for input and output of magnetic field (see Fig. 2b), which leads to appearance of the additional second hysteresis $\Delta R(H)$. These hysteretic curves for the field output for $H > H_X$ located lower than the curves for the field input that corresponds to a lower value of the sample magnetization for the field output. This does not correspond to $M(H)$ behavior. $M(H)$ dependences have only one hysteresis for $H < H_A$, wherein $H_A$ is the field of anisotropy blocking reorientation of the magnetic moments of the particles, which results in residual magnetization. Such type of MR hysteresis we have previously observed [5-7,11]. It relates to the percolation nature of tunneling conductance of granular system at low temperatures, and is due to switching of a small number of current channels at the field input and field output. The second unusual feature of negative MR behavior is the reduction of $\Delta R(H)$ with increasing magnetic field, starting with relatively small fields $H_{max} \cong 0.2$

T, where $H_{max}$ is field of maximum of the negative MR. This effect was diminished and gradually disappeared with rising the temperature or the measuring current (Fig. 3).

For a solid solution $Cr_{1-x}Fe_xO_2$ reduction of negative MR was replaced by its re-growth, after some field $H_A$, equal to the field of convergence for ascending and descending branches of $R(H)$ dependence (Fig. 2a). By analogy with the behavior of $M(H)$ hysteresis we called $H_A$ the field of anisotropy. When $H < H_A$ the system is in the blocked state in which the inequality $\tau > t_i$ is fulfilled, where $\tau$ - the relaxation time for the spin system, $t_i$ - the time of observation. From there it follows that for $H < H_A$ we should see the dependence of MR hysteresis loop on the rate of magnetic field grows $dH/dt$. We measured the $R(H)$ dependences for Sample 2 at different rates of magnetic field grows. The results of these measurements are presented in Fig. 4. It is seen that the shape of MR curves is actually determined by the rate of the magnetic field changing $dH/dt$. The changing of dH/dt varies the ratio between the time of observation and the spin relaxation time of the system. So the results shown in Fig. 4 actually mean that the magnitude and character of the $\Delta R(H)$ dependence is determined by the rate of relaxation of the magnetic spin subsystem. From Fig. 4, it is seen that the decrease of $dH/dt$ reduces the coercive field $H_p$, the maximum achievable value of the MR, the field at which the negative MR reaches its maximum ($H_{max}$) and the area of the hysteresis loop. As we can see, the maximum values of negative magnetoresistance in weak fields can only be achieved at high rates of the magnetic field changing and correspond nonequilibrium state of the magnetic subsystem. At the minimum rate $dH/dt = 0.0029$ T•s$^{-1}$, the system approaches to equilibrium ($\tau < t_i$), and the extremum on the RM curve at the field input degenerate into a small shoulder at $H_0$. The reduction of MR with magnetic field input in this case is not observed (Fig. 4b). However, even at a minimum rate $dH/dt = 0.0029$ T•s$^{-1}$ the equilibrium state of the magnetic system is not achieved. It can be seen from Fig. 5, which shows the changing of $\Delta R(H_{fixed})$ with time at different fixed fields. The measurement protocol was as follows. After the sample magnetization reversal, we gradually introduced the magnetic field for 40 seconds with a minimum rate equal to 0.0029 T•s$^{-1}$, up to the value 0.0029 T•s$^{-1}$ * 40 = 0.116 T. After that, the field was stabilized and we continue recording the time dependence $\Delta R(t)$ at $H$ = 0.116 T. Gradually increasing the field with increments $\Delta H$ = 0.116 T, this procedure was repeated until it reaches an equilibrium state of the system. In this state the reduction of $\Delta R(H)$ at $H$ = const has no longer been observed. From the data in Fig. 5 one can see that as the field $H$ approaches $H_A \cong 0.6$ T, the rate of the change of MR with time (d$\Delta R$ ($H$)/d$t$) gradually decreases. Results of these timebase experiments indicate that in addition to the external magnetic field and the dipole field of the sample the spin system is influenced by the demagnetization factor that prevents magnetization of the sample at the magnetic field input and tends to bring the system back to equilibrium when input of the magnetic field is stopped. At low temperatures, the conductivity of the granular system has a percolation character. Relaxation processes during the magnetization reversal can be related to the influence of the anisotropy fields and to the interaction of the magnetic moments of the ferromagnetic particles forming conducting chains, with the magnetic moments of the rest of the sample (i.e. with the magnetic moments of the neighboring particles that are not participate in conductivity). In the needle-like particles with a diameter 24 ÷ 34 nm and a length of 300 nm, there is a uniaxial magnetic anisotropy, which enhances the coercive force and prevents the magnetization of the sample. During compression of the samples consisting of such particles the magnetic texture usually forms. The powder particles during compaction preferably stacked in the plane perpendicular to the axis of pressing, being randomly distributed in a plane. Because in these particles the easy magnetization axis approximately coincides with the axis of symmetry of particle, while removal or stabilization of the magnetic field the magnetic moments of the particles tends to rotate in the direction of the easy axis and the negative magnetoresistance is reduced as we see in Fig. 5.

The above mentioned peculiarities of the behavior of $\Delta R(H)$ are seen only at sufficiently low temperatures, when the MR is determined by the conductivity of a small number of "optimal" chains of granules with maximum probability of tunneling. This means that the processes of magnetization reversal of a macroscopic ensemble of ferromagnetic granules and quasi one-dimensional chains of granules (low-dimensional structures) occur by different scenarios. In conditions of activation conductivity the number of conducting channels continuously decreases with decreasing temperature, and at a sufficiently low temperature, a percolation grid can be reduced to a single conducting channel [12]. These «Optimal» conducting chains may have few weak bonds (high-resistance tunnel junctions) with high activation energy. These high-resistance contacts will determine the overall conductivity of the system. At a fixed temperature the increase in the magnetic field lowers the potential barriers. As a result the negative magnetoresistance increases sharply in small fields. Because the energy levels of the electronic states in the neighboring granules are always blurred, the conduction electrons must acquire some energy (e.g. from phonons), before the tunneling occurs. At low temperatures the number of phonons with the necessary energy is limited, so in a single tunnel junction the number of carriers entering the tunnel contact may be greater than the number of carriers leaving the contact at the same time period $\Delta t$. As a result, when you input the magnetic field, the charge transfer process is blocked at a fairly small field and the conductivity decreases (negative magnetoresistance is reduced). When the rate of the magnetic field input decreases, part of the electrons have enough time to gain energy from phonons and tunnel, and during the time $t \leq \tau$ the magnetic system relaxes to the equilibrium state. This reduces the peak of MR at $H_{max}$ and leads to the gradual disappearance of the second hysteresis. However, this process is accompanied by a general lowering of the tunnel MR.

Thus, the common features of low-temperature behavior of the tunnel MR in pressed $CrO_2$ powders can be explained by the granular structure and by the formation of the magnetic texture. At the same time, solely on the basis of our results, we can not make an unambiguous conclusion about the mechanism of spin relaxation in the samples studied. Further consideration of the problem requires additional experiments.

FIGURE CAPTIONS

Fig.1. Temperature dependence of resistivity for Sample 1 (open circles) and Sample 2 (closed circles). Measurement current $J$ = 100 μA.

Fig.2. a) MR hysteretic curves taken at T = 4.25 K in a magnetic field $\mathbf{H} \parallel \mathbf{J}$ (d$H$/d$t \cong$ 0.021 T/s). Open circles - Sample 1. Dark circles - Sample 2.

　　b) The appropriate MR dependences on a larger scale at low fields' area. $H_P$ – the field that corresponds to the maximum of the positive magnetoresistance. $H_X$ – the field of crossing of the branches for the field input and the field output dependences.

Fig.3. a) Dependences $\Delta R(H)/R(0)$ for Sample 2, recorded at different temperatures ($J$ = 100 μA, $\mathbf{H} \parallel \mathbf{J}$).

　　b) Dependences $\Delta R(H)/R(0)$ for Sample 2, recorded at different currents $J$ = 200 μA, 2000 μA, 5000μA, 10000 μA ($T$ = 4.93 K, $\mathbf{H} \parallel \mathbf{J}$). The rate of the magnetic field input d$H$/d$t$ = 0.021 T/s.

Fig.4. a) MR hysteresis curves for Sample 2 at $T$ = 4.25 K, written at different rates of the magnetic field pulling: − - 0.25 T/s, ,- 0.125 T/s, ∀ - 0.037 T/s. $\mathbf{H} \perp \mathbf{J}$, $J$ = 100 μA.

　　b) MR hysteresis curves for Sample 2 at $T$ = 4.25 K, written at the minimal rate of the magnetic field pulling d$H$/d$t$ = 0.0029 T/s.

Fig.5. Reduction of the negative magnetoresistance in the Sample 2 with time at different fields $\mathbf{H} \perp \mathbf{J}$. The rate of the magnetic field input d$H$/d$t$ = 0.0029 T/s.

Table. Basic characteristics for the samples studied.

$H_c$ - coercive force, $M_{max}$, $M_{res}$ - the maximum and residual magnetization of the samples, respectively, $K_s$ – squareness of the hysteresis loop, **a**, and **c** - rutile-type lattice parameters, $V_{cell}$ – volume of the unit cell, $S_{sp}$ - specific surface, $d_{ef}$ – the effective diameter of the particle. Magnetic characteristics correspond to the temperature T = 293 K.

| Sample number | $H_c$, Oe | $M$, A·м²/kG | | $k_s$ | a, Å | c, Å | $V_{cell}$, Å³ | $S_{sp}$ m²/g | $d_{ef}$ nm |
|---|---|---|---|---|---|---|---|---|---|
| | | max | res | | | | | | |
| 1 | 522 | 83.9 | 36.6 | 0.437 | 4.4253 | 2.9120 | 57.0265 | 34 | 24 |
| 2 | 761 | 75.3 | 34.6 | 0.459 | 4.4270 | 2.9140 | 57.1095 | 24 | 34 |

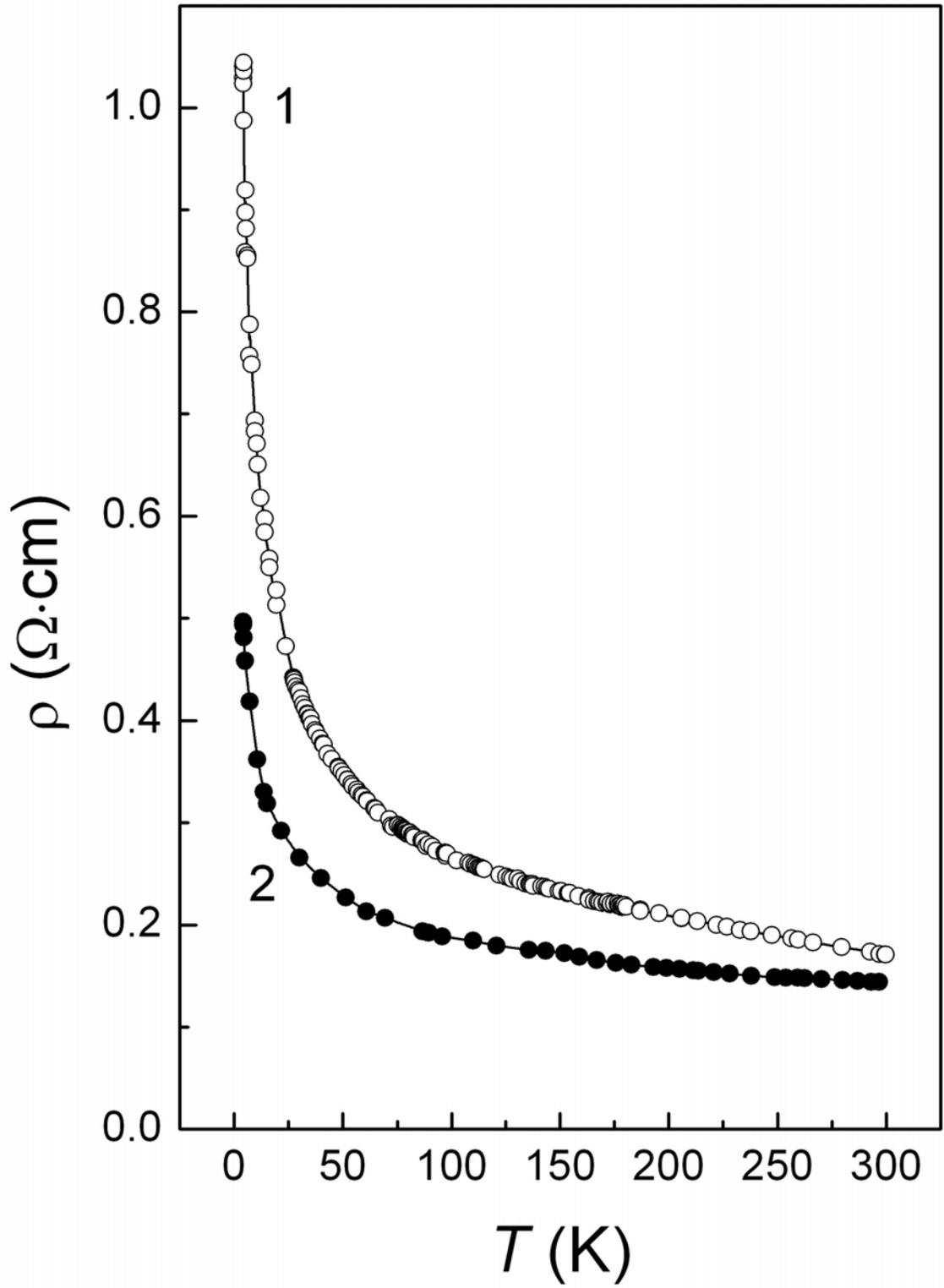

Fig. 1.

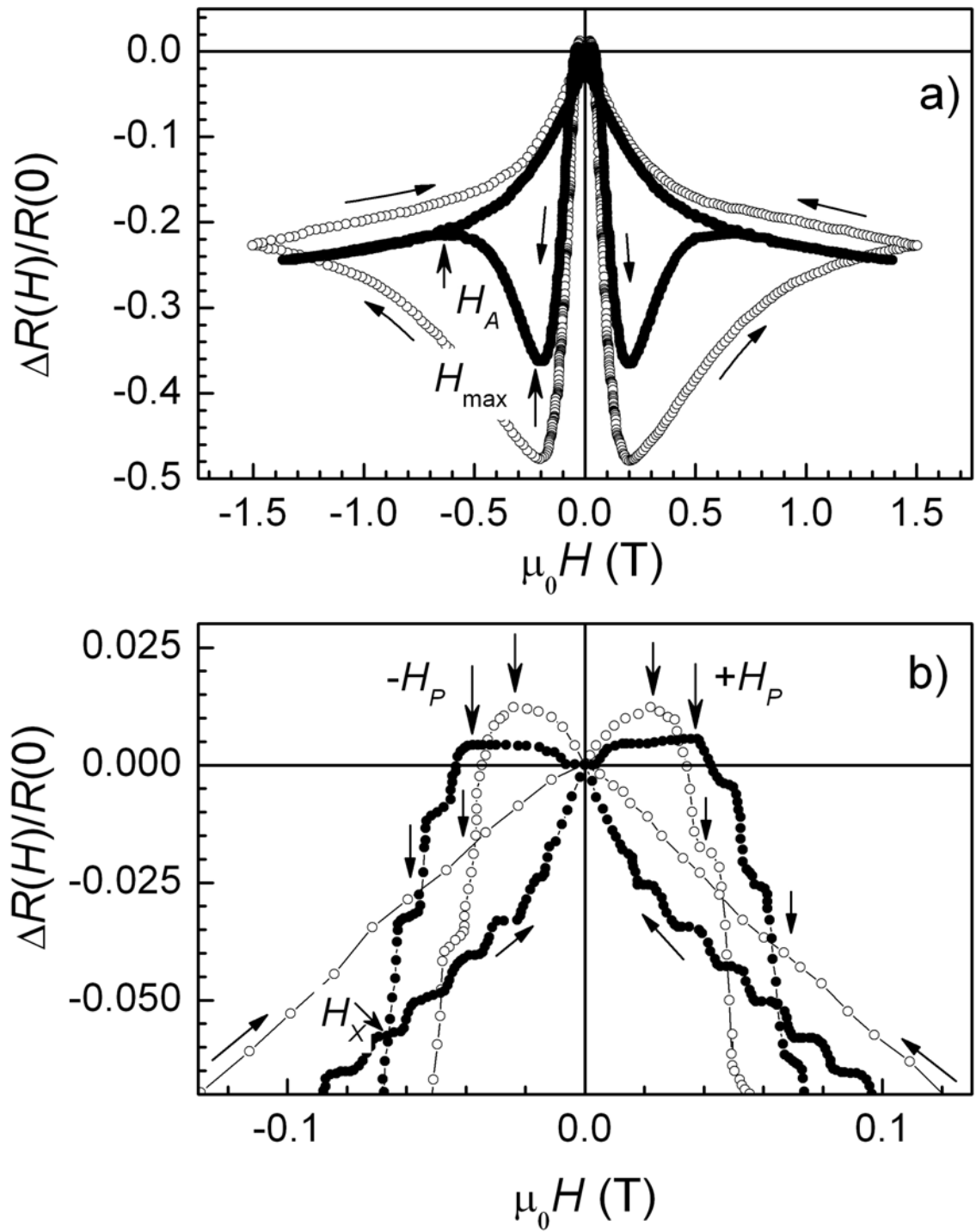

Fig. 2.

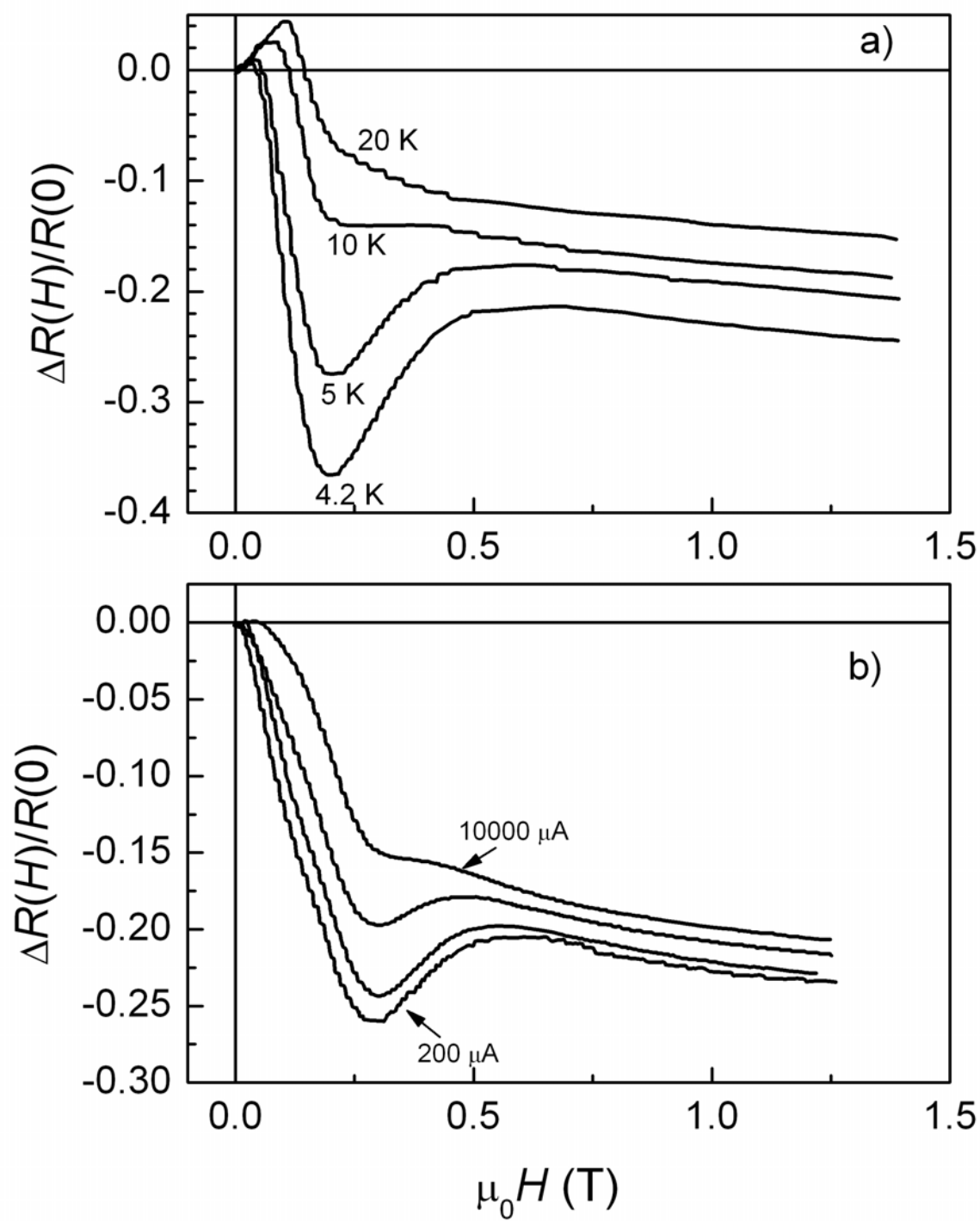

Fig. 3.

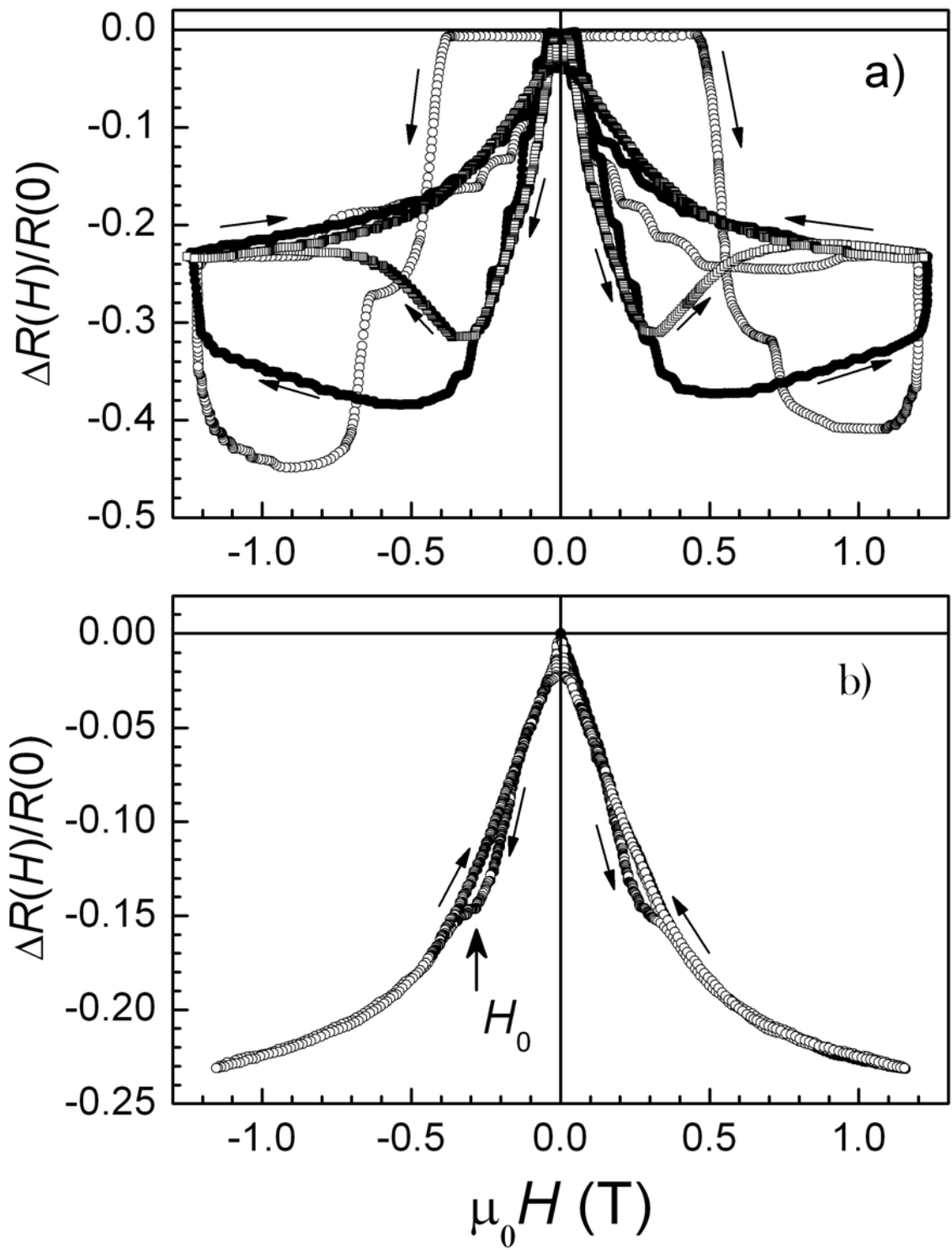

Fig. 4.

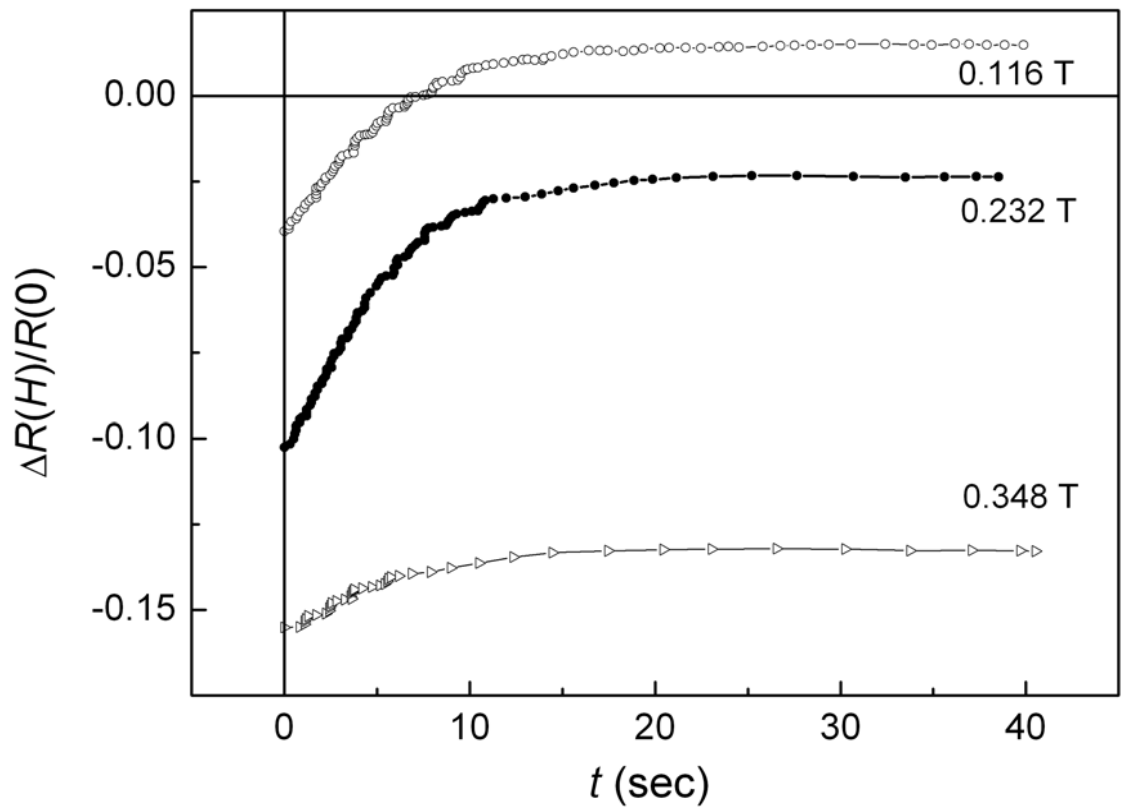

Fig. 5.

This text further to be published in printed version …